\documentclass[letterpaper]{article} 
\usepackage{aaai25}  
\usepackage{times}  
\usepackage{helvet}  
\usepackage{courier}  
\usepackage[hyphens]{url}  
\usepackage{graphicx} 
\usepackage{natbib}  
\usepackage{caption} 
\usepackage{algorithm}
\usepackage{algorithmic}

\usepackage{xcolor}
\newcommand{\answerYes}[1]{\textcolor{blue}{#1}} 
\newcommand{\answerNo}[1]{\textcolor{teal}{#1}} 
\newcommand{\answerNA}[1]{\textcolor{gray}{#1}} 
 
\usepackage{newfloat}
\usepackage{listings}
\DeclareCaptionStyle{ruled}{labelfont=normalfont,labelsep=colon,strut=off} 
\floatstyle{ruled}
\newfloat{listing}{tb}{lst}{}
\floatname{listing}{Listing}
\usepackage{booktabs}
\newcommand{\numNotes}{2,750,917}
\newcommand{\numNotesSample}{1,869,517}
\newcommand{\modelcoveragepct}{0.894}
\newcommand{\NMRprop}{0.871}
\newcommand{\CRHprop}{0.094}

\newcommand{\firstnotes}{134,525}
\newcommand{\nreauthor}{96,269}

\title{Threats to the sustainability of Community Notes on X}
\author {
    Zahra Arjmandi-Lari\textsuperscript{\rm 1}\thanks{The authors assert a Creative Commons Attribution (CC BY 4.0) License for this preprint},
    Alexios Mantzarlis\textsuperscript{\rm 2},
    Tom Stafford\textsuperscript{\rm 3},
}

\affiliations{
    \textsuperscript{\rm 1}Inria, Bordeaux, FR\\
    \textsuperscript{\rm 2}Independent Researcher\\
    \textsuperscript{\rm 3}University of Sheffield, Sheffield, UK\\
    z.arjmandi@gmail.com, mantzarlis@protonmail.com, t.stafford@sheffield.ac.uk

}

\begin{document}

\maketitle

\begin{abstract}
Community Notes are emerging as an important option for content moderation. The Community Notes system pioneered by Twitter, now known as X, uses a bridging algorithm to identify user-generated context with upvotes across political divides, supposedly spinning consensual gold from partisan straw. It is important to understand the nature of the community behind Community Notes, especially as the feature has now been imitated by several billion-user platforms. We look for signs of stability and disruption in the X Community Notes community and interrogate the motivations which may be driving users to contribute. We show substantial churn in the population of users who author Notes. We note that the majority of notes considered “helpful” (and therefore shown to users on X) is low, ~10\%, and declining. We conduct a novel analysis of the impact of having a note published, which requires being considered “helpful” by the bridging algorithm, utilising a regression discontinuity design. This allows stronger causal inference than conventional methods used with observational data. Our analysis shows than any positive effect on future note authoring of having a note published is unclear. This analysis has implications for the future of Community Notes on X and the extension of this approach to other platforms.
\end{abstract}

%

\section{Introduction}
Content moderation is an essential function of a digital platform, yet it is also a highly disputed one \cite{gillespie2018a}. Top-down decisions to remove or label (or not) a piece of content have led to advertiser boycotts, consumer complaints, and political pressure. While large online platforms have extensive policy guidelines describing violative behavior, the diversity of the speech they govern makes it nearly impossible to apply those policies completely consistently, and their scale means that even a small percentage of errors can affect thousands or millions of users.

Moderation decisions can be particularly challenging when it comes to content that is harder to define, such as misinformation. There is at least a baseline of consensus around what constitutes sexual or violent content; misinformation is by definition contextual and requires high-quality contradictory evidence to be available. Still, the public \cite{ejaz2024people} and regulators \cite{husovec2024} expect action from platforms in this realm, and unchecked online misinformation risks negatively affecting outcomes from health behaviors \cite{allen2024quantifying} to political polarization \cite{budak2024}.

\begin{figure}[ht]
\centering
\includegraphics[width=0.75\columnwidth]{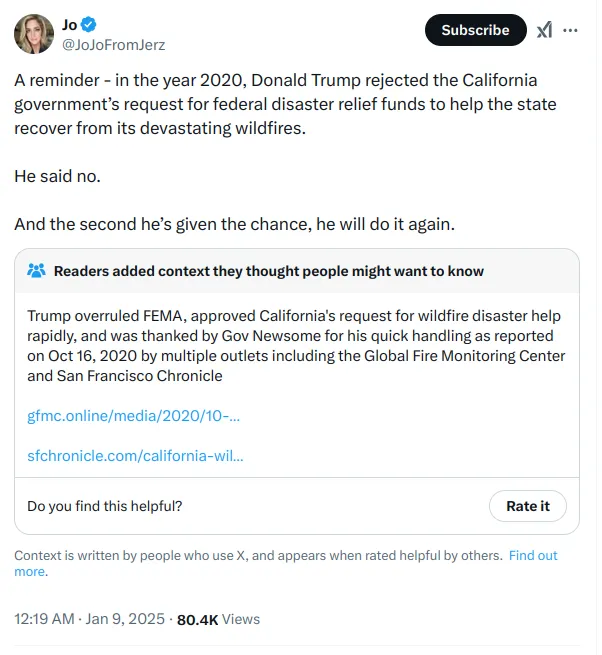}   
\caption{An example of a post on X.com, with a Community Notes appended under the label ``Readers added context they thought people might want to know"). Screenshot from \url{https://x.com/JoJoFromJerz/status/1877148134972739607} on 2025-01-09
}
\label{CNexample}
\end{figure}

This context informed the decision by X (then Twitter) to launch Birdwatch in January 2021 \cite{twitter2021,wojcik2022birdwatch}. Users were invited to provide context to posts they thought were misleading in the form of “notes” that other users could rate as helpful and not helpful (Figure \ref{CNexample}). Crucially, the decision on which Notes to publish was devolved to an algorithms which identified Notes which still receive endorsements once the effects of political polarisation are statistically discounted. The algorithms belong to a family of consensus-identifying algorithms known as `bridging algorithms' (they stand in contrast to the `engagement algorithms' which are central to most mainstream social media platforms, which identify content which receives the most engagement, regardless of where it comes from). Only those notes that are deemed helpful by a sufficiently diverse set of raters are affixed to the offending tweet and seen by all of the platform’s users.

1,408,000 users contributed or rated at least one note to the feature, now called Community Notes, by June 2026 (up from 1280,000 by December 2025 and 930,000 by December 2024),.

It has also served as a more or less explicit inspiration for similar features launched on Meta \cite{testing_CN_Meta}, TikTok \cite{TikTok}, and YouTube \cite{youtube_test}.

\section{Prior work}
Previous research has explored the efficacy of crowd-sourced efforts in detecting and researching misinformation. \citet{martel2024} concluded that small groups of diverse participants are generally effective at identifying misinformation, with precision rates similar to those of professional fact checkers. \citet{karadzhov2024effect} found that greater cognitive diversity was associated with better problem solving in a cognitive task. \citet{zhao2023} found that crowd-checking and traditional fact-checking projects in Taiwan largely agreed on their ratings. Agreement between Birdwatch data and the corpus of fact-checking articles carrying ClaimReview metadata was also significantly greater than disagreement \cite{saeed2022}. 

Community Notes' fact checks can be of high quality. Medical professionals who rated 205 COVID-19-related notes found that 97\% of them were accurate \cite{CN_Covid}. An important caveat is that Community Notes are frequently reliant on professional fact-checkers \cite{borenstein2025replaceFactCheckers} and other sources of information like Wikipedia and news media \cite{solovev2025references}. In addition, the Community Notes system may be vulnerable to manipulation or bias \cite{truong2025vulnerable}. Examples would be the finding that claims which are more effortful to research, define and refute are less likely to be rated \cite{wack2026laziness}, as well as results suggesting that the system may underdeliver precise on polarising topics \cite{bouchaud2026undermoderate}, or for topics with more than one dimension of polarisation \cite{andreou2026faultlines}.

As with other fact-checking interventions, Community Notes appear to have a moderately positive effect on the spread of misinformation, including by reducing its reach and inducing authors of false posts to delete them \cite{chuai2024community, renault2024collaboratively}, though this effect may be too slow to have a decisive impact \cite{chuai2024did}. A survey conducted by Twitter itself found that helpful notes reduced the likelihood that a user agreed with the flagged tweet by about 26\% \cite{wojcik2022birdwatch}.

In addition to the impact of the published Notes, many Community Notes users author Notes annotated as "Note Not Needed". These Notes have their sole intended audience as other Community Notes users, not the wider X platform, and thus reflect the repurposing of the Community Notes system as a venue for debate in and of itself ~\cite{razuvayevskaya2025timeliness}.

Evidence that Community Notes and other crowd-sourced fact-checking systems can be successful makes it more urgent to address their scalability and sustainability. We need to understand the mechanisms that drive the function of Community Notes and may support or hinder the long-term viability of Community Notes and similar systems, especially as they are more widely adopted.

On this front, we know relatively little about why users join Community Notes and what motivates them to contribute corrective information that is more likely than not never to be seen, for a private platform owned by a mercurial billionaire. The filtering effect of X’s scoring algorithm that takes into account “not only how many contributors rated a note as helpful or unhelpful, but also whether people who rated it seem to come from different perspectives”\footnote{https://communitynotes.x.com/guide/en/under-the-hood/ranking-notes}. This means only about 10\% of all notes being attached to tweets; the vast majority never get past the Community Notes interface (see Results, Figure \ref{fraction_helpful}.)\footnote{verifiable via data at https://notetracker.socialmediadata.org/}

\citet{allen2022birds} have found that counterpartisanship can be a driving factor, as X users are motivated to correct someone they disagree with. \citet{yoon2025} conducted interviews with eight contributors to crowd-sourced fact-checking efforts to understand the support they seek from each other. Beyond these valuable insights, however, little is known about what keeps contributors engaged in Community Notes. Our work builds on \citet{pilarski2024community}'s, focusing on what Community Notes contributors choose to fact-check by trying to answer why they continue to contribute to the program.

\subsection{Causal inference in observational data}

Observational data places limits on causal inference: it is usually not clear what causes what \cite{pearl2018book}. Although digital platforms have made available large, naturally occurring datasets which promise insight into psychological and behavioural function \cite{goldstone2016,stafford2017testing}, statistically controlling for confounding constructs is harder than often assumed \cite{westfall2016statistically,young2011}. Where direct experimentation is ethically or practically unfeasible, it is possible to use a family of quasi-experimental methods which, under when assumptions hold, allow causal inference \cite{huntington2021effect}. Ultimately these methods rely on a foundation of causal graph analysis \cite{klebel2026}. While there is not space for a full exposition here, this current report deploys and explains a single one of this family of quasi-experimental methods, the regression discontinuity design \cite[RDD;][]{cattaneoRDD}.

\subsection{Research questions}

To better understand the community behind Community Notes we first describe the changes in the size and nature of the community over recent years, focusing particularly on the authorship of notes (as distinct from mere rating of notes). We analyze the rate at which new authors join and leave the system. This provides the charactertisation of the Community Notes system on X which is essential for further understanding.

Our research question concerns the sustainability of the Community Notes system. Is the rate at which new authors join the system sufficient? How long do new authors stay? What factors affect their retention?

On this final question, we provide an analysis of the incentive power of having notes published.

\section{Data collection and methods}
At the time of writing, the data for Community Notes on X is open and available to any registered user\footnote{At https://x.com/i/communitynotes/download-data and licensed for use under the X Developer Agreement and Policy}. The code is also openly published\footnote{https://github.com/twitter/communitynotes and licensed under an Apache-2.0 license}. This transparency is a deliberate element of the Community Notes system \cite{making_CN}. 

The data includes a “note” dataset, containing all notes, their ID, note author ID, respective tweet ID, creation time, classification (i.e., “misinformed or potentially misleading”, “not misleading”), and their text (called “summary”). There are also “rating”,  “user enrollment”, and “note status history” datasets, containing the activity of users and the history of notes’ status (i.e., “needs more rating (NMR)”, “currently rated helpful (CRH)”, and “currently rated not helpful (CRNH)”). 

By combining the Community Notes algorithm (from the code repository) with the data (i.e. ratings from users, linked to Note and tweet IDs), it is possible to independently recalculate Note scores. Due to the size of the data this process is facilitated by access to High Performance Computing (HPC) resources.

Our analysis uses the published data from Community Notes. We used the algorithm without modification to recalculate scores for individual Notes. The code contains multiple variations of the bridging algorithm. Where relevant, we restrict our analysis to the ~89.4\% of Notes published due to scores derived from the core algorithm and three major variants. 

In calculating the effects of having a published note on future note-writing, we exclude all “No Notes Needed” (NNN) notes. These are a specific type of note that allows contributors to argue against appending a note to a specific post. Though these notes can be scored for helpfulness, we make the assumption that, because these Notes are not published by the system, these NNN contributions are not motivated by publication of the Note but rather by the effect of the note on the overall publication of other Notes.

Our analysis starts with the full population of \numNotes{} Notes authored within the Community Notes system from its inception to June 1st 2026.
Our analysis code, which generates the statistics and figures presented here, is available at \url{https://github.com/zahra-arjm/community_notes}.

\section{Results}

\subsection{A growing community driven by a subset of power users}
\begin{figure}[ht]
\centering
\includegraphics[width=0.95\columnwidth]{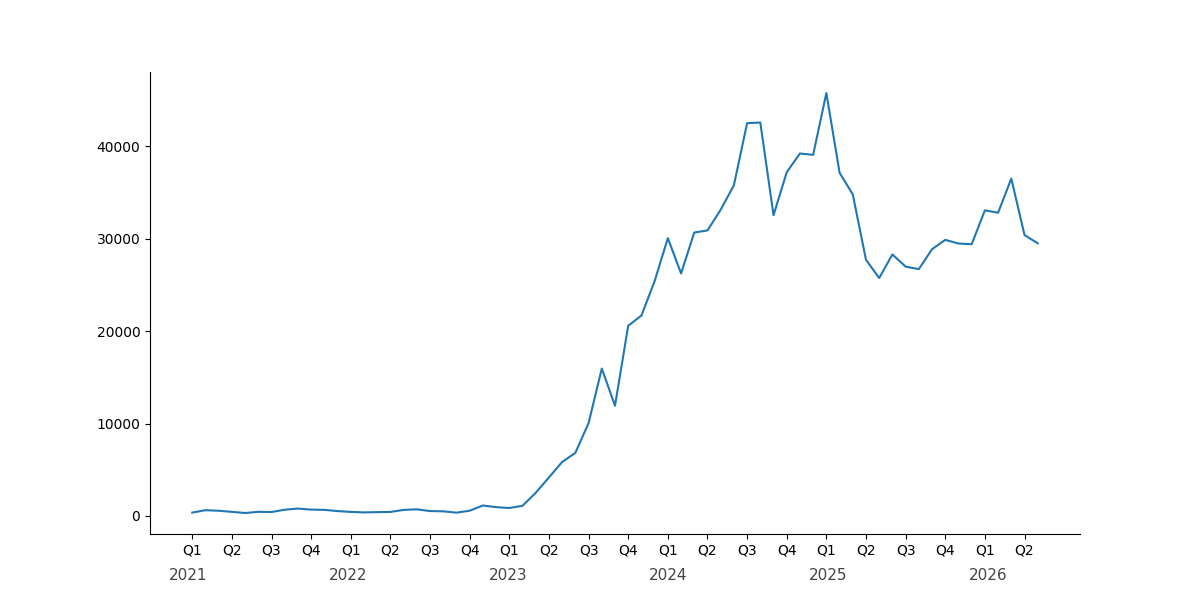}   
\caption{Monthly Active Authors (MAAs) contributing at least one note during the reference month}
\label{monthly_authors}
\end{figure}

\subsubsection{Monthly active users surged as the program expanded and stabilized in 2024 (Figure \ref{monthly_authors}).} 

Community Notes launched in January 2021, and spent most of the ensuing two years with fewer than 1,000 monthly active authors (MAAs) – which we define as users who contributed at least one note that month, regardless of whether it was rated helpful. Following Elon Musk’s takeover of Twitter in Oct. 2022, the platform ramped up access to the feature\footnote{https://x.com/CommunityNotes/status/1578004575990202370}. Over the course of 2023, many new countries were onboarded, and by the end of the year the program counted more than 20,000 MAAs. 

By the time Community Notes was admitting users from the whole world in the second half of 2024\footnote{https://x.com/CommunityNotes/status/1839035926963695858}, the user base had stabilized at around 40,000 MAAs. (Including all users who have rated at least one note over the past month, the active user base goes up to 605,000.) For comparison, Wikipedia had more than 250,000 monthly active editors at the time of writing (2026), similarly defined as users who edited at least one page in the previous 30 days \cite{wiki_stat}. Because of this significant variance in user numbers over time, we choose to focus much of our analysis on the period starting in January 2024 (comprising \numNotesSample{} Notes).

\begin{figure}[ht]
\centering
\includegraphics[width=0.95\columnwidth]{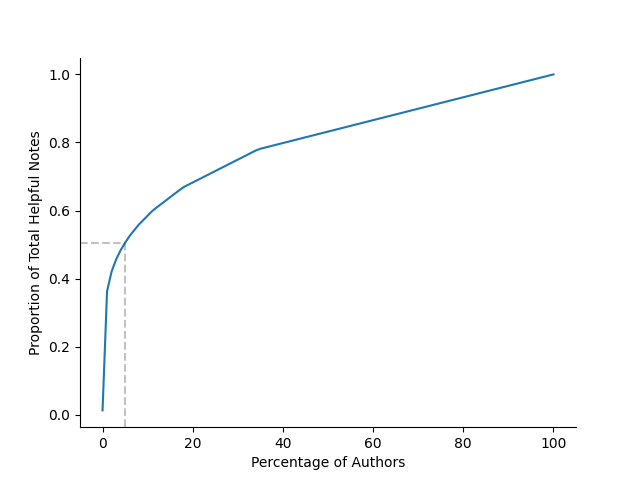}   
\caption{Share of helpful notes authored by percentile of authors since 2024. Dashed line shows median (50th percentile) point.}
\label{cum_contributions}
\end{figure}

\subsubsection{Most helpful notes are written by a minority of contributors (Figure \ref{cum_contributions}).} 

As with other crowd-sourced efforts, a small share of contributors contributed far more than the median user. Since 2024, the top 1\% of Community Notes authors wrote 36\% of all helpful notes, and the top 7.5\% were responsible for 50\% of all helpful notes.

\subsection{Community Notes authors persist and get replaced, but are not all equal}
\begin{figure}[ht]
\centering
\includegraphics[width=0.95\columnwidth]{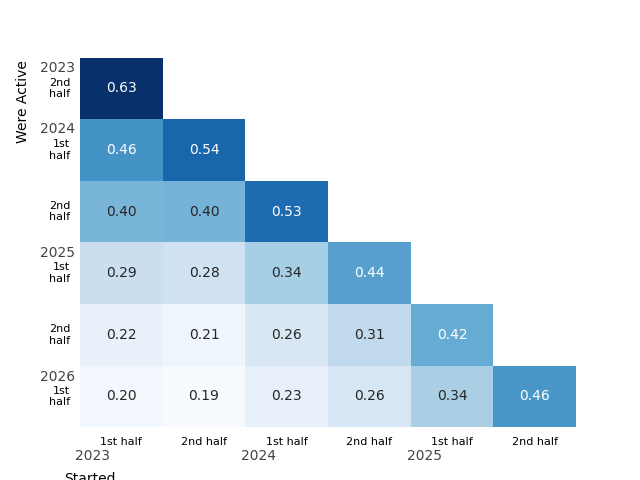}   
\caption{Persistence rate of authors based on the time of their first note.}
\label{heatmap}
\end{figure}

\subsubsection{Community Notes authors have a significant churn rate (Figure \ref{heatmap}).}

A large group of Community Notes authors do not stay active past their first note. Drop-out from first-time authors from the same period can be read from Figure \ref{heatmap}) top-to-bottom along the same column. Fewer than half (46\%) of the authors who contributed to Community Notes in the first half of 2023 were still active one year later, and only 29\% by the first half of 2025. The year-on-year persistence of authors appears to have worsened. This can be read from Figure \ref{heatmap}) by looking along the diagonals, which show the retention for cohorts of authors who started at different times, at comparable time-removes. 40\% of first-time authors from the second half of 2023 are still contributing one year later in the in the second half of 2024, while the comparable one-year-late rate for those who joined in first half of 2024 is 34\%, and 31\% for those who joined in the second half of 2024.

\begin{figure}[ht]
\centering
\includegraphics[width=0.95\columnwidth]{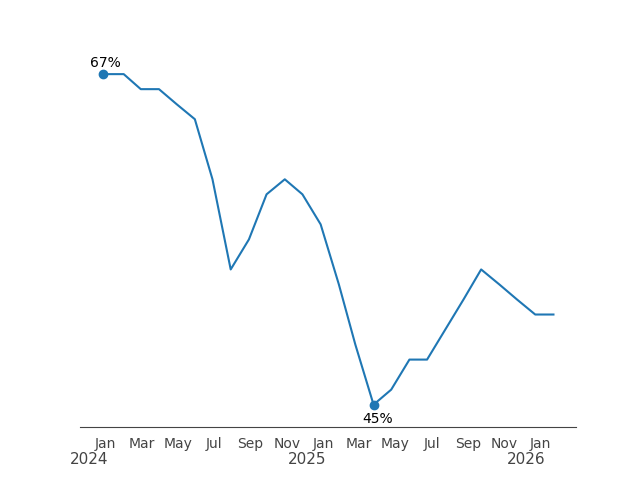}   
\caption{Fraction of authors who stayed active (defined as authoring another note within 4 months of their first note).}
\label{fraction_active_after_1st}
\end{figure}

\subsubsection{Author retention rate is declining but still healthy (Figure \ref{fraction_active_after_1st}).}

Even as many authors choose to drop out of the program, Community Notes has been able to draw new authors in at a higher rate than those it loses. That is, in part, explainable by the program’s rapid expansion in 2023-2024. From 2024, with the system covering users of the platform from all regions, that rate of expansion ceased to be possible.
To better understand the dynamics of author retention, we calculate the percentage of new authors who go on to post a second note in the 4 months after their first note. By assessing ongoing activity using a rolling 4-month window, we are able to treat all time points equally (avoiding the issue that time points closer to the end of the series have a reduced future window in which to see author activity). The change in this proportion over time is a key indicator of the program’s sustainability, since it is sensitive to the on-boarding pipeline by which new authors join and cement their participation in the Community Notes system. The proportion has been, on average, decreasing (Figure \ref{fraction_active_after_1st}). We suggest that this proportion is a leading indicator of Community Notes’ capacity to maintain a healthy community of active contributors.

\subsection{Most notes are unseen, which may have a long-term effect}

\begin{figure}[ht]
\centering
\includegraphics[width=0.95\columnwidth]{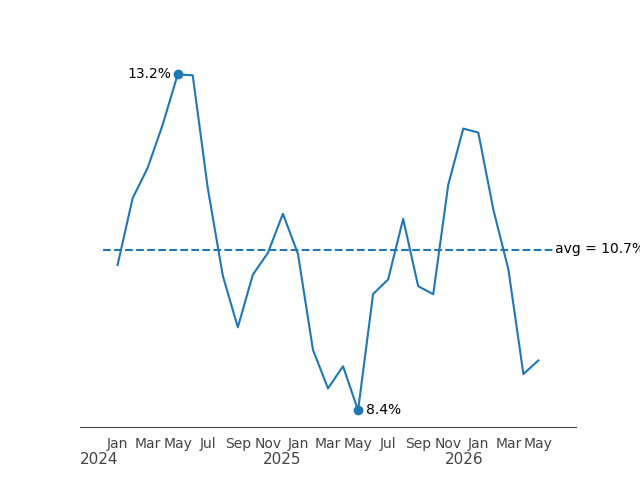}   
\caption{Percentage of helpful notes.}
\label{fraction_helpful}
\end{figure}

\subsubsection{Helpful notes are declining as a share of the total (Figure \ref{fraction_helpful}).} 

In the long-term, fulfilling the proposition of the Community Notes program requires notes to be both authored and to be endorsed as helpful, and so published for general users to see. The proportion of authored notes which are labeled helpful by the bridging algorithm has been broadly declining since May 2024.

The Community Notes system calculates a score for note helpfulness. It is not sufficient to have a large number of users rate a Note helpful for it to have a high helpfulness score. As defined by the bridging algorithm at the heart of the Community Notes system, a Note must also receive positive ratings from a sufficiently diverse set of users, such that these ratings are not predicted by those users’ partisan tendencies (for more on the algorithm, see \citealt{What_do_I_think_about_CN, understanding_CN_algorithm}). The low proportion of Notes published means that most authored Notes are not seen by general users, and most authors will experience the majority of their Notes failing to get published. This itself may have consequences for author retention.

\subsection{Analysis of effect on author retention of having a note published}


The bridging algorithm reconciles ratings from users in the Community Notes system to produce a helpfulness score. Above a value of 0.4 a Note is considered helpful and so it is eligible to be published. 

The 0.4 threshold establishes a discontinuity, which may allow insight into the effect that having a Note published has on author retention with the Community Notes system. A discontinuity such as in the publication threshold is one circumstance in which stronger causal inference may be possible from observational data, using a formal analysis called regression discontinuity design \cite[RDD;][]{cattaneoRDD}. A regression discontinuity design  contrasts “near hits” with “near misses”, based on a threshold where small differences generate different outcomes.  This comparison allows causal inference on the effect of the outcome which is conditional on crossing the threshold. The reasoning is that for cases sufficiently close to the threshold it is only random noise which determines whether an individual case is just above or just below, and so such cases can be fairly compared. In our case, we compare first-time Community Notes authors whose first note is just above or below the threshold score for publication on X. Our outcome variable is the probability of an author going on to write subsequent notes. The overall design will address the causal effect of an author seeing their first Note published, compared to not. More loosely, the analysis speaks to the motivational effect having Notes published on author retention.

\subsubsection{Our sample}

We restrict our analysis to authors who publish their first Note between 2024-01-01 and 2026-06-01, for Notes whose final publication status was decided by the original bridging algorithm (`core model'), as well as an `expansion model', and two main variants of each (full details are in the supplementary material). Together these four models drive the publication decision for \fpeval{round(100*\modelcoveragepct{},1)}\% of Notes in our sample. Of these we only consider notes with $>0$ ratings and which are for original posts classified by the Note author as `MISLEADING'. This produces a sample of \firstnotes{} Notes by first-time authors, of which \nreauthor{} went on to write at least one more Note.

Because the X Community Note algorithm and rating data are open it is possible to reconstruct the helpfulness scores for each Note in the data. While it is technically feasible to reconstruct the helpfulness score at each point of time since the system inception, this requires more compute by several orders of magnitude than we have access to, so we calculate final helpfulness scores of each Note given all the ratings received. 

\begin{figure}[ht]
\centering
\includegraphics[width=0.95\columnwidth]{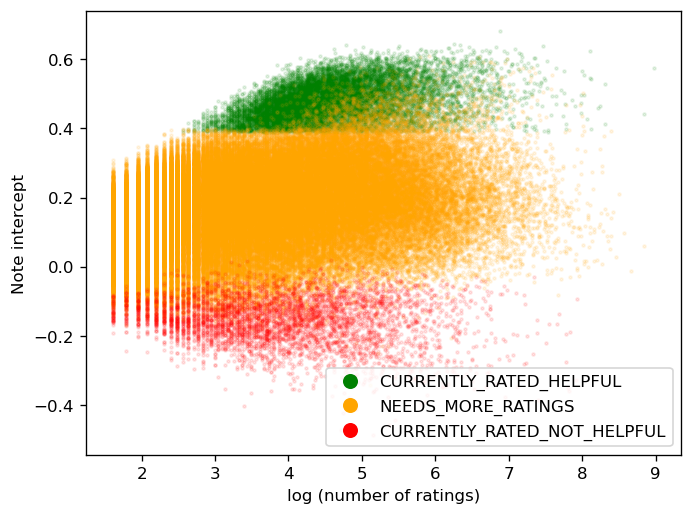}   
\caption{All-time Note helpfulness score according to the standard algorithm and three major variants against log Number of user ratings, colored by current note status (rated “helpful": green; rated “not helpful": red; rated “needs more rating”: orange), for Notes authored since 2024-01-01.}
\label{score_vs_numRating}
\end{figure}

Figure \ref{score_vs_numRating} shows our recalculated helpfulness scores against the log number of ratings, and coloured according to recorded final note status. This confirms that the vast majority of notes do not receive a sufficient diversity of ratings to be rated Helpful or Not Helpful (\fpeval{round(100*\CRHprop{},1)}\% of all Notes are labeled `Currently Rated Helpful', with \fpeval{round(100*\NMRprop{},1)}\% labeled `Needs More Ratings', with the remainder 'currently Rated Not Helpful'). This plot shows the clear discontinuity at the score threshold for a Note being rated helpful: only Notes with scores above 0.4 result in a Note being rated helpful and shown to users. Scores can decrease as well as increase with additional ratings, which is why some Notes rated `Need more ratings' (in orange) are above the 0.4 threshold. No Notes with scores below 0.4 have a final not status ('Currently Rated Helpful').

\subsubsection{Regression Discontinuity Design}

It is possible to conduct a regression discontinuity design (RDD) analysis using simple linear (OLS) regression, and it is helpful to consider this as the foundation for understanding the broad class of analyses. Two regression lines are fitted, one for data above the threshold and one for data below the threshold, bounded by some bandwidth to prevent including points so far away from the threshold that they are not comparable (for example, very low quality notes that would never be close to achieving publication will be written a different class of author than those who write Notes at threshold for publication). In our case, Note helpfulness is the predictor (running variable). The outcome variable is future note publication by that author (0 or 1). The two regression lines make it possible to estimate an effect size at the the point of discontinuity in the running variable, the local average treatment effect (LATE). In out case, this is an estimate of the causal impact of getting published, for those Notes at the threshold.

The basic OLS method has some negative features, for example instability in the regression slopes and intercepts due to points further from the threshold, and the need to declare an arbitrary bandwidth size for data inclusion. For these reasons the basic OLS method is dispreferred to methods which use more robust estimation methods, using data-derived bandwidth selection \cite{calonico2014robust, calonico2018effect}. For implementation, we rely on the \textit{rdrobust} package \cite{calonico2014rdrobust}, with the default options. The package also provides a range of additional options, including built-in sensitivity tests. The full specification is available in the online supplementary material, along with the code for reproducing all results.

\subsubsection{Robustness}

\begin{figure}[ht]
\centering
\includegraphics[width=0.95\columnwidth]{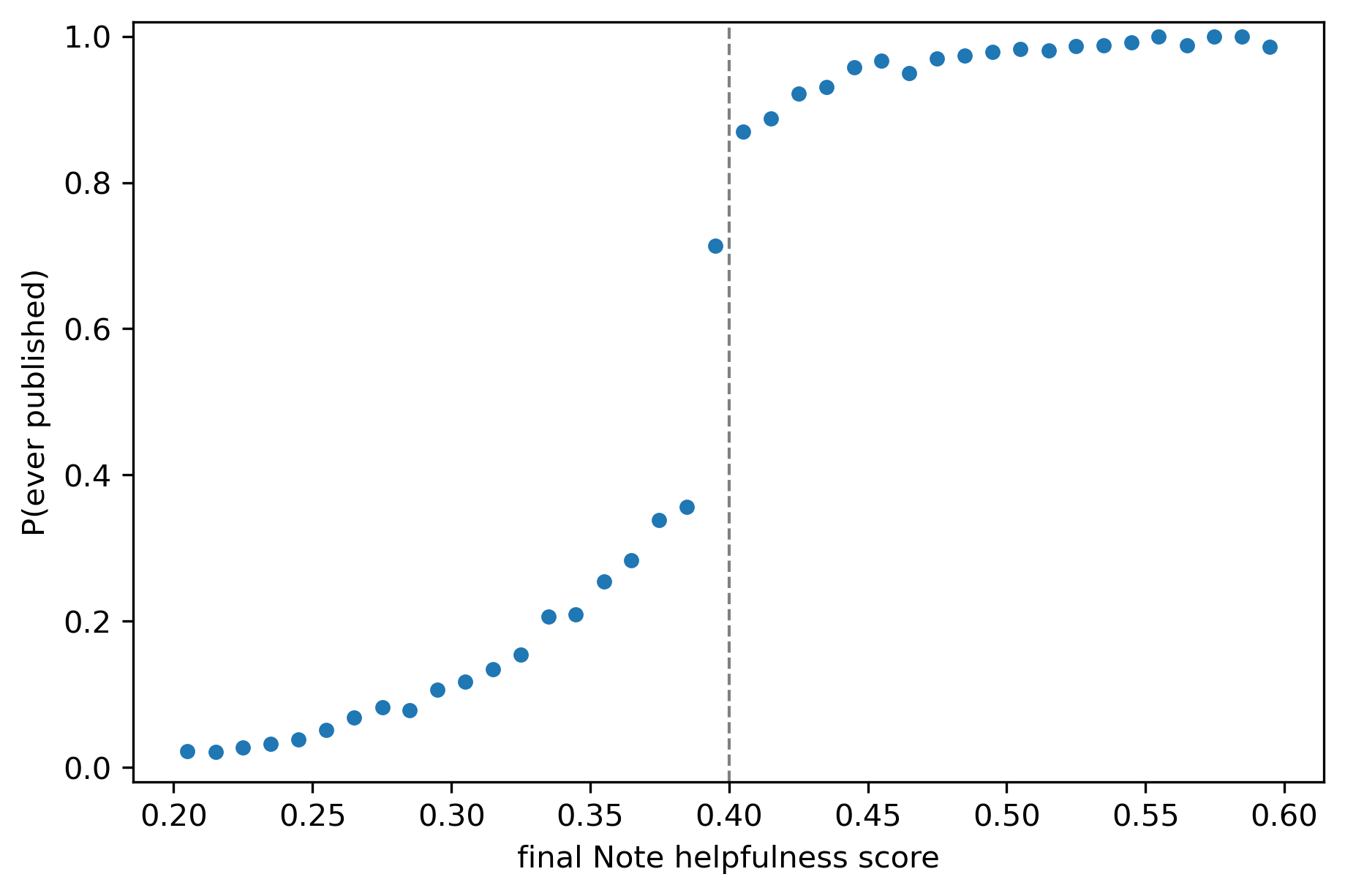}   
\caption{Probability of Notes ever having been published, binned by final Note helpfulness score.}
\label{ever-published}
\end{figure}

The final helpfulness scores for each Note do not completely reflect the full publication history of each Note, since Notes can increase and decrease in helpfulness as ratings accrue. The Community Note data includes some indicators of a Note ever having been published, which we can aggregate (see supplement for details) to construct an `ever published' variable. This variable that confirms that the there is a discontinuity around the 0.4 threshold, but the probability of a Note being having ever been published does not flip from 0 to 1 at this threshold, instead increasing from a floor level of about 3\% for Notes with helpfulness around 0.2, to 100\% for Notes with helpfulness around 0.6. See Figure \ref{ever-published}. For intermediate final Helpfulness scores the probability of ever having been published rises steadily, with a jump so that Notes in the range 0.35 - 0.38 have a probability of ever having been published of less than 50\% and Notes in the range 0.40 - 0.42 have a probability of ever having been published of more than 90\%. This non-sharp discontinuity is handled by using a fuzzy RDD design, a model will takes the `ever published' status of a Note as an mediating variable.

This inspection of the probability of a Note ever being published revealed an anomalous region; Notes with a helpfulness score of 0.39 - 0.40 had an probability of ever having been published of 77.9\%. An explanation for this can be found in documentation for the Community Notes system which introduces a tolerance for Notes to drop to 0.39 (but no more) and retain their publication status. To exclude these anomalous data points we conduct our analysis only on Notes with helpfulness scores which are at least 0.01 from the threshold (a so called 'donut', since it removes the middle from the window which the RDD covers).

\begin{figure}[ht]
\centering
\includegraphics[width=0.95\columnwidth]{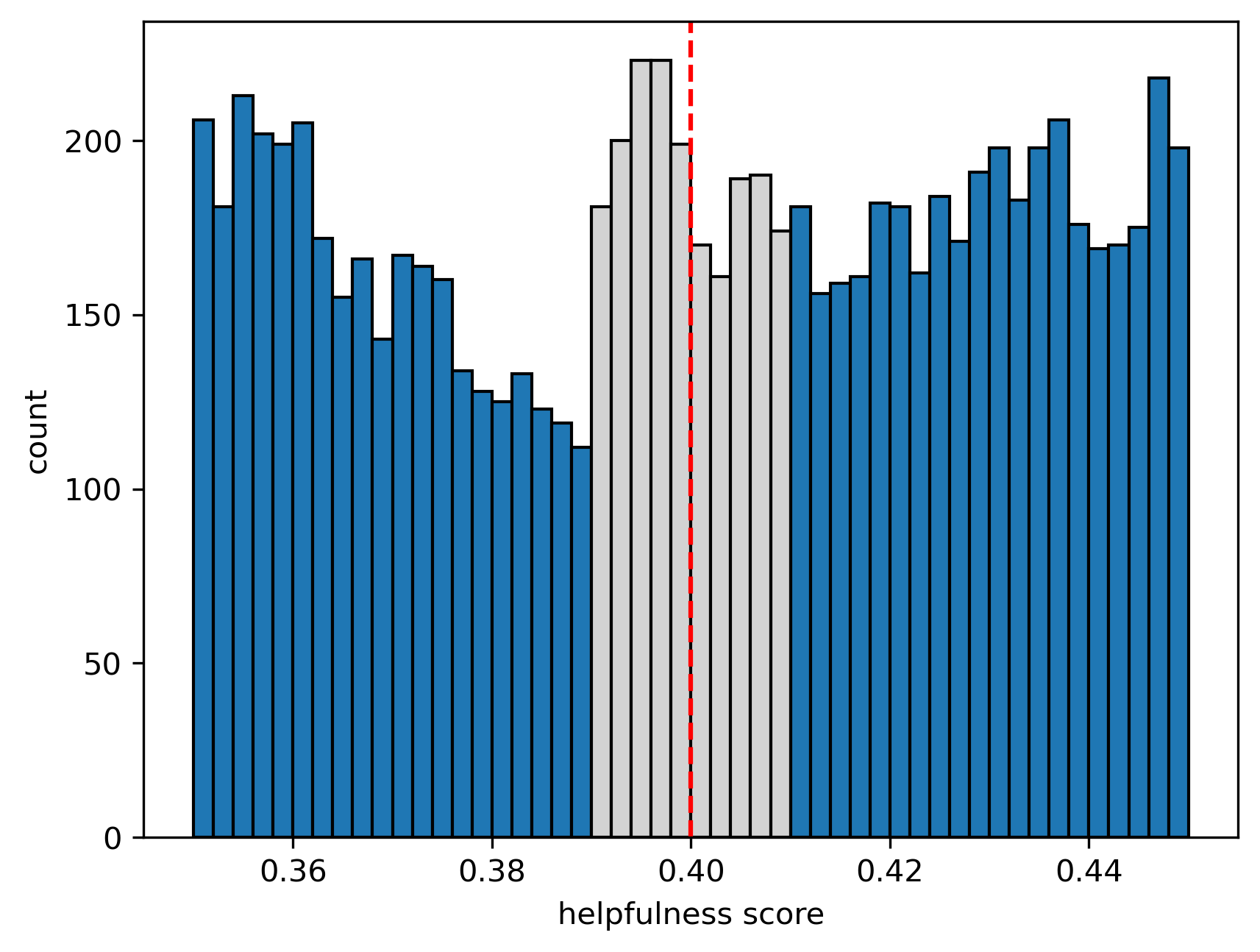}   
\caption{Histogram of count of Notes within 0.05 of the 0.4 helpfulness threshold. Those Notes in the grey bins covered in the donut exclusion.}
\label{densityplot}
\end{figure}

A standard robustness check for RDD is to inspect the counts of the points within the analysis window. This is shown in Figure \ref{densityplot}. This histogram shows that the anomalous zone of 0.39 - 0.40, where Notes are below the threshold but more likely to be ever have been published, is also anomalous in the count of Notes. 

\subsubsection{Result}

The prior sections established that while there is no sharp discontinuity in our proposed outcome (an author having their first note published), the data are sufficient to support a fuzzy RDD analysis, where the outcome is a probabilistic function of a known intermediate variable (ever-having-been-published).

The RD robust method identifies an MSE-optimal bandwidth of $h = 0.065$ and provides an estimated treatment effect of $\hat{\tau} = +0.096$. This reflects a +9.6 percentage point increase in the probability of writing a second note if an author's first note is published (conventional local-linear estimator; robust bias-corrected 95\% CI $[-3.3, +23.5]$, $p = 0.14$; Calonico, Cattaneo, \& Titiunik, 2014).

The rate of republishing for notes which fall below the cutoff is $\sim 69.2\%$. The treatment effect of 9.6\,pp implies a 13.9\% relative lift over the counterfactual baseline; however, this difference is non-significant.

\begin{table*}[htbp]
\centering
\caption{Main fuzzy-RDD estimate.}
\label{tab:tab-main}
\begin{tabular}{rrrlr}
\toprule
$h$ & $\hat\tau$ & $p$ (robust) & 95\% CI (robust) & N in h \\
\midrule
0.065 & 0.096 & 0.140 & [-0.033, 0.235] & 9999 \\
\bottomrule
\end{tabular}
\end{table*}

\begin{figure}[ht]
\centering
\includegraphics[width=0.95\columnwidth]{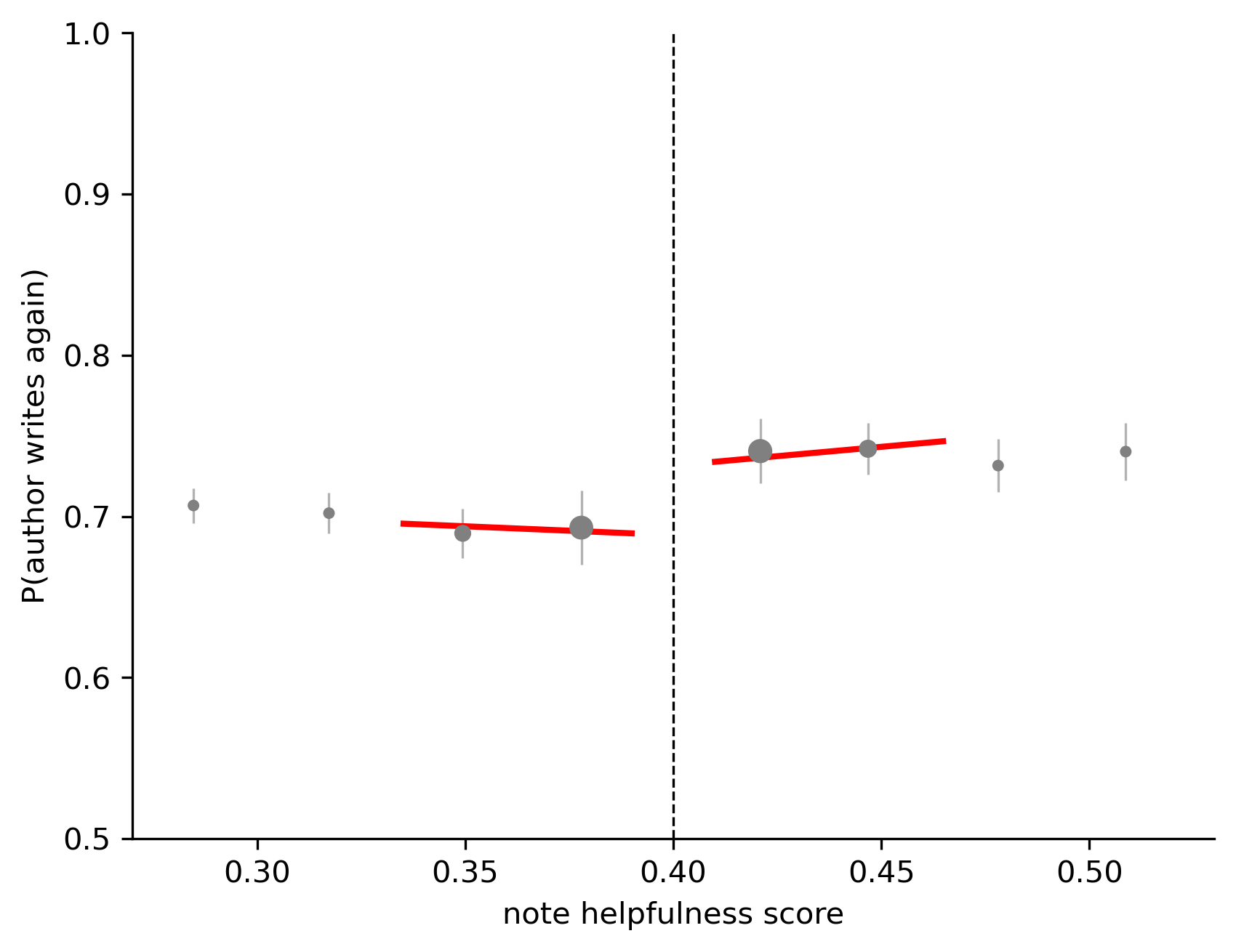}   
\caption{Probability of an author writing a second note dependent on the helpfulness score of their first note (total N = 105,375). Red lines are local linear fits for points each side of the 0.4 publication threshold, using observations within the MSE-optimal bandwidth (h = 0.065). Grey points are the bin means (with 95\% confidence intervals); point size is proportional to each bin's kernel weight. Notes within 0.01 of the threshold are excluded.}
\label{RDD}
\end{figure}

Figure \ref{RDD} visualizes this result, demonstrating a numerical discontinuity at the threshold; a mean vertical shift before/after (left/right) can be seen.

This is the first time, to our knowledge, that stronger methods of causal inference have been applied to the analysis of notes, and it suggests that Community Notes authors may be demotivated by the declining rate of Note publication.

\subsubsection{Sensitivity checks}

While the result presented is the most appropriate and robust according to the authors' belief, it is possible to further interrogate the result by testing the sensitivity of the underlying effect to alternative specifications.

An alternative, simpler, model is the OLS approach. Taking the data-derived bandwidth from the rdrobust method, we can run a RDD using OLS, including the donut exclusion justified above.

A bandwith of 0.064 identifies 4931 treated Notes and 4880 control Notes. The OLS estimated treatment effect at the cutoff is 0.0325, $p = 0.192$, $t(df) = 1.304(9807)$

This confirms a numerically positive, but non-statistically significant, effect.

It is also possible to re-run the rdrobust model but with different cutoff thresholds. Since there is no discontinuity at thresholds of 0.30, 0.35 or 0.45, we should expect the model to find no large or significant effect of crossing these thresholds (contrawise, if a model reports large significant effects at these thresholds we should suspect either our model or our understanding of the data). For consistency, we keep the donut exclusion of +/- 0.01 around the threshold, but omit the fuzzy RDD specification using the `ever published' variable (since this fails to align with other thresholds it does not make sense to adjust estimates of effect size by it). The results are shown in table \ref{tab:placebo-rdd}.

\begin{table*}[htbp]
\centering
\caption{sharp RDD, donut of 0.1, with thresholds of 0.30, 0.35, 0.40 and 0.45}
\label{tab:placebo-rdd}
\begin{tabular}{lrrrlr}
\toprule
Cutoff & $h$ & $\hat\tau$ & $p$ (robust) & 95\% CI (robust) & N in h \\
\midrule
0.30 & 0.077 & -0.020 & 0.393 & [-0.057, 0.022] & 23747 \\
0.35 & 0.081 & -0.019 & 0.298 & [-0.074, 0.023] & 18605 \\
0.40 & 0.087 & 0.050 & 0.073 & [-0.004, 0.103] & 15148 \\
0.45 & 0.042 & 0.032 & 0.387 & [-0.054, 0.138] & 5641 \\
\bottomrule
\end{tabular}
\end{table*}

We note that the true threshold produces the largest effect estimate ($\hat{\tau}$). The 0.30 and 0.35 threshold suggest small negative effects, the 0.45 threshold suggest a small positive effect. None of the four effects are statistically significant (n.b. the result for the 0.4 threshold is not identical to the result already reported because this is a sharp RDD, not a fuzzy RDD; it omits the ever-published variable). Only the 0.4 threshold effect is close to statistical significance.




\section{Discussion}


X’s Community Notes has been successful at maintaining an active contributor base since 2024, but it is vulnerable. The period up to 2024 was marked by continual expansion of the Community Notes system to X users in new locales. Now that expansion of eligible contributors is no longer feasible – all X users are eligible to join Community Notes – the incentives and churn of the existing community becomes a more important issue.

We show that most notes are produced by a minority of contributors, and that contributors do churn out of the program. The fraction of authors who remain in the program year-on-year has declined since 2023, suggesting a reduced capacity to hold on to authors.

Crucially, a large and growing majority of notes are not published. This may affect the program in the long term.

We used a Regression Discontinuity Design (RDD) to investigate the causal effect of a first-time author having their first Note published. To our knowledge, this is the first time a RDD analysis has been applied to the Community Notes system. Our estimate of the effect was non-significant, although numerically positive (i.e. in the expected direction if Note publication encouraged subsequent authoring). This result is compatible with Note publication having no \textit{immediate} effect on retention, or it having an effect which is below the statistical resolution of the current data. The result is also compatible with retention-effects over the longer term (e.g. authors are more likely to continue authoring if some of their Notes are published).

Our RDD analysis suggests that getting a first Note published is not a significant factor in determining users' future likelihood of contributing to Community Notes. This finding is echoed in observational data of a period of downtime in the program in May 2025\footnote{https://x.com/CommunityNotes/status/1927112800960176547}.  During a short period of time when notes were not visible on X at all, contributors dropped to 57\% of their usual range, but - notably - they did not cease completely. It is unclear whether this is because users didn’t understand that the program had vanished or didn’t care, but either way, it suggests that while seeing note publication may affects propensity to contribute, it cannot be the sole determinate, at least in the short term.

Although others have highlighted the role counterpartisanship plays in the motivation of Community Notes authors \cite{allen2022birds}, if Notes, motivated by counterpartisanship or not, are not published, authors may be funnel their efforts elsewhere.

All in all, these results speak to risks to the sustainability of the Community Notes system. Author retention and publication are declining. The motivations for authoring are evidently complex, extending, beyond publication. Without understanding them, the Community Notes system relies on declining momentum.


During the preparation of this report, X announced their integration of AI-authored notes into Community Notes\footnote{https://x.com/communitynotes/status/1940132205486915917}. It is not clear if this is a response to a perceived decline in human participation, nor is it clear how competition with AI-authored notes will affect human Note authors. What is evident in the data, however, is that a handful of AI Note Writers contribute the majority of all Helpful notes [CITE: https://indicator.media/ai-community-notes-tracker], while human contributions to the published corpus remain far below the historical peaks.

In an aligned move, when the X platform introduced the Grok AI there was a decrease in teh engagement with the Community Notes system, suggesting that it provided an competing method of engaging with fact-checking on the platform \cite{zhou2026notes}.


We have focused on Note authoring, not Note rating, which is a second, essential, activity of Community Note participants. Note publication may also affect authors' participation via rating, as well as their retention as Note authors. Investigating this might allow a more comprehensive view of participation in the Community Note system and hence the dynamics of motivation and retention.

Another limitation of this work is that it deals with the behaviour of Community Note participants, trying to infer their motivations, but not speaking to them direct. Due to anonymity of system users, it is not possible to directly approach users, but future work could seek to engage with Community Users who voluntarily identify and so investigate motivations for participation more directly.

While our analysis focused on the likelihood that a note gets published, future work might explore what incentivizes repeat contributors whose notes never get published.

An obvious analytic approach to the current analysis would be compute the contemporaneous helpfulness score of Notes, and so be able to conduct a sharp RDD, rather than relying on the final helpfulness score, as we do here. This is beyond the computational resources we can access.

Ongoing changes to the Community Note system also pose hard-to-quantify threats to the sustainability of the system on X and on other platforms where it is deployed. AI Note authors and the integration of AI fact-checking into platforms (as in the case of Grok) may impact Community Note participation. Future work could look at willingness to contribute by different subsets of the Community Notes user base, segmented by frequency of publication or rating, or by geography.


The Community Note system is a promising innovation, crowdsourcing the intelligence of a platform's own users to support content moderation. It is clear that as well as relying on the wider ecosystem of journalism and fact-checking, the Community Note system will also need to rely on community of users on a platform. The vulnerabilities to author recruitment and retention suggest that platforms may need to do more in the future to support at community.

\section{Acknowledgments} 
Thanks to Nemanja Vaci for the discussion of natality-mortality dynamics, and to Andreas Vlachos, and Sudhamshu Hosamane for advice.

\bibliography{aaai25}

\end{document}